\documentclass[preprintnumbers,article,amsmath,amssymb,floatfix,9pt,prd,
superscriptaddress,nofootinbib]{revtex4}
\usepackage[colorlinks=true, pdfstartview=FitV, linkcolor=blue, citecolor=red, urlcolor=magenta]{hyperref}
\usepackage{bbm}
\usepackage{amsfonts}
\usepackage{dsfont}
\usepackage{mathrsfs}
\usepackage{latexsym}
\usepackage{epsfig}
\usepackage{epstopdf}
\usepackage{epstopdf}
\usepackage{graphicx}
\usepackage{amssymb}
\usepackage{amsmath}
\usepackage{dcolumn}
\usepackage{caption,subcaption}
\usepackage{bm}
\usepackage{multirow}
\usepackage{color}
\usepackage{comment}
\usepackage{xcolor}
\usepackage{ragged2e}

\begin{document}

\title{\textbf{Probing Realistic Neutron Star Equations of State via the Chameleon Mechanism}}

\author{Debadri Bhattacharjee}
\email{debadriwork@gmail.com}
\affiliation{IUCAA Centre for Astronomy Research and Development (ICARD), Department of Physics, Cooch Behar Panchanan Barma University, Vivekananda Street, District: Cooch Behar, Pin: 736101, West Bengal, India.}

\author{Grigorios Panotopoulos}
\email{grigorios.panotopoulos@ufrontera.cl}
\affiliation{Departamento de Ciencias F\'isicas, Universidad de La Frontera, Casilla 54-D, 4811186, Temuco, Chile.}

\author{Pradip Kumar Chattopadhyay}
\email{pkc\_76@rediffmail.com}
\affiliation{IUCAA Centre for Astronomy Research and Development (ICARD), Department of Physics, Cooch Behar Panchanan Barma University, Vivekananda Street, District: Cooch Behar, Pin: 736101, West Bengal, India.}

\begin{abstract}
We investigate the impact of the Chameleon mechanism on structural properties of neutron stars adopting realistic nucleonic equations of state. The scalar–matter coupling parameter $\beta$ controls the departure from the corresponding general-relativistic configurations, affecting both the maximum mass and the stellar radius. Our numerical results demonstrate that the chameleon mechanism can have a significant impact on the macroscopic properties of neutron stars, even when realistic equations of state are employed. Moreover, well established constraints on masses and radii of neutron stars put stringent limits on the allowed numerical values of the coupling.
\end{abstract}

\maketitle

\date{\today}

\section{Introduction}\label{sec1}

Neutron stars (NSs) \cite{NS1,NS2,NS3} provide a unique environment to investigate gravity under extreme physical conditions. Formed following the gravitational collapse of massive stars, these compact objects exhibit central densities comparable to or exceeding nuclear saturation density, strong gravitational fields, and highly relativistic matter. Their macroscopic properties, including mass, radius, compactness, and tidal deformability, depend sensitively on both the equation of state (EOS) of dense matter and the underlying theory of gravity. General relativity (GR) \cite{GR} has successfully described a wide range of observations involving compact objects; nevertheless, the possibility of deviations from Einsteinian gravity on astrophysical and cosmological scales continues to motivate the investigation of alternative gravitational frameworks, see for instance \cite{mod1,mod2,mod3,mod4,mod5,mod6,mod7}. In particular, neutron stars offer a promising setting in which to examine whether additional gravitational degrees of freedom can produce measurable changes in stellar structure \cite{P1,P2,P3}.

\smallskip

Among the proposed extensions of GR, scalar–tensor theories, with the Brans-Dicke model \cite{Brans} being the archetype one, are of particular interest because they introduce a scalar field that can modify the effective gravitational interaction between matter and spacetime. A central challenge for such theories is to reconcile potentially significant modifications of gravity on large scales with the stringent constraints imposed by local gravitational experiments. Screening mechanisms provide a possible solution to this problem by suppressing the observable effects of additional fields in environments where deviations from GR are tightly constrained. The chameleon mechanism, introduced by Khoury and Weltman \cite{P4,P5}, achieves this suppression through a density-dependent effective mass of the scalar field. As a consequence, the field may become short-ranged in high-density environments while retaining a longer interaction range in low-density regions. This environmental dependence has made chameleon models an important framework for investigating modifications of gravity in both cosmology and astrophysics.

\smallskip

The essential feature of the chameleon mechanism is the nonlinear interaction between the scalar-field potential and its coupling to matter. The effective potential depends on the local matter density, causing the scalar field to evolve toward a density-dependent minimum. Around sufficiently dense objects, the resulting effective mass can become large, suppressing the field-mediated force and allowing the predictions of GR to be approximately recovered. However, screening should not automatically be assumed to eliminate all observable effects in compact stars. The relativistic pressure, strong curvature, and nontrivial radial variation of matter properties can influence the scalar-field configuration and its coupling to the stellar fluid. Consequently, the structure and stability of neutron stars in chameleon gravity must be examined through a self-consistent treatment of the scalar field and the relativistic matter distribution \cite{P2,P3}.

\smallskip

The equilibrium structure of a nonrotating, spherically symmetric neutron star in GR is governed by the Tolman–Oppenheimer–Volkoff (TOV) equations \cite{P6,P7}, which combine hydrostatic equilibrium with the Einstein field equations and an appropriate EOS. In scalar–tensor gravity, the additional scalar degree of freedom modifies the gravitational field equations and, depending on the chosen conformal frame and matter-coupling prescription, introduces additional contributions to the hydrostatic balance and stellar geometry. The resulting system of modified equilibrium equations must therefore be solved together with the scalar-field equation and the EOS. Regularity at the stellar center, vanishing pressure at the stellar surface, and suitable asymptotic conditions on the scalar field provide essential boundary conditions for constructing physically meaningful stellar configurations. The resulting solutions permit a systematic comparison between predictions from GR and those obtained in the presence of chameleon screening.

\smallskip

Previous investigations have demonstrated that screening mechanisms can influence compact-star observables in ways that depend on both the gravitational model and the properties of dense matter. Brax, Davis, and Jha \cite{P2}, for example, studied neutron stars in chameleon and environmentally dependent dilaton models, finding modifications to the mass–radius relation and identifying a significant degeneracy between gravitational parameters and the stellar EOS. Dima, Bezares, and Barausse \cite{P3} subsequently investigated the dynamical behaviour of chameleon neutron stars, examining radial oscillations, stability, and scalar radiation during gravitational collapse. Their results indicate that configurations exhibiting effective screening can remain stable against sufficiently small radial perturbations, although their oscillation spectra may differ from the corresponding GR predictions. These studies establish that neutron stars can serve not only as probes of equilibrium modifications to gravity but also as laboratories for testing its dynamical consequences.

\smallskip

Recent work has further emphasized the importance of studying the structural response of compact stars to screened modifications of gravity. In particular, Bachs-Esteban, Lopes, and Rubio \cite{P8} investigated the impact of the chameleon mechanism on white dwarfs, demonstrating that the scalar field can alter the pressure gradient and shift the predicted mass–radius relation relative to the standard gravitational description. Although white dwarfs and neutron stars occupy different density and compactness regimes, this study highlights the broader relevance of stellar structure calculations for constraining scalar–tensor theories. For neutron stars, the relativistic nature of the matter and the possible importance of the pressure contribution to the trace of the energy–momentum tensor introduce additional complications. In particular, regions in which the pressure becomes sufficiently large relative to the energy density may alter the effective scalar-field dynamics and affect the degree of screening within the stellar interior \cite{P2,P3}.

\smallskip

From an observational perspective, the study of neutron-star structure has become increasingly important with the development of complementary gravitational-wave and X-ray measurements. Gravitational-wave observations of binary neutron-star systems constrain tidal deformability, while X-ray observations provide independent information about stellar masses and radii \cite{P9}. These observables depend on the internal density distribution and the gravitational field and can therefore be used to test alternative theories of gravity. Nevertheless, distinguishing modifications of gravity from uncertainties in the EOS remains a substantial challenge. A systematic investigation of the chameleon mechanism must consequently consider the joint dependence of stellar observables on the scalar-field parameters, the matter coupling, and the adopted description of dense nuclear matter.

\smallskip

In this context, the present study investigates the structural properties of neutron stars within the chameleon mechanism, focusing on the scalar-field configuration and its influence on relativistic stellar equilibrium. By solving the corresponding coupled field and hydrostatic equations, the analysis aims to determine how the chameleon parameters modify the stellar pressure and density profiles, gravitational mass, radius, and mass–radius relation. Particular attention is given to the comparison with GR predictions and to the dependence of the resulting configurations on the assumed EOS. This approach provides a framework for identifying the parameter regimes in which screening effectively suppresses deviations from GR and those in which potentially significant structural effects remain. Ultimately, understanding these effects is essential for assessing the viability of chameleon gravity in the strong-field regime and for determining whether neutron-star observations can provide meaningful constraints on screened scalar–tensor theories.

\smallskip

The remainder of this paper is organised as follows. In Section II, we present the chameleon gravity framework and review the modified Tolman–Oppenheimer–Volkoff equations governing static, spherically symmetric neutron-star configurations. Section III introduces the realistic BSk24 and SLy4 equations of state and describes the construction of the corresponding mass–radius relations. In Section IV, we present and discuss our numerical results, examining the dependence of the maximum mass and radius on the scalar–matter coupling parameter and deriving constraints from massive-pulsar observations and neutron-star radius measurements. In Section V we discuss scalar field normalisation and boundary conditions, while Section VI is devoted to the Harrison-Zel'dovich-Novikov stability criterion. Finally, Section VII summarises our main findings and discusses possible directions for future work.

\section{Tolman-Oppenheimer-Volkoff equations in chameleon theory}\label{sec2}

The chameleon theory \cite{P4,P5}, featuring a screening mechanism, is expressed through the following action functional \cite{P8}:
\begin{equation}
     \mathfrak{S}=\int d^{4}x\,\sqrt{-g}\,\Bigg[\frac{R}{16\pi}-\frac{1}{2}\nabla_{i}\phi\nabla^{i}\phi-V(\phi)\Bigg]+\mathcal{L_{M}}\Bigg[\Psi_{m};A^{2}(\phi)g_{ij}\Bigg], \label{eq1}
\end{equation}
where we have used the relativistic system of units, $G=1$ and $c=1$. In Eq.~(\ref{eq1}), $g$ represents the determinant of the fundamental metric in the Einstein frame, $R$ is the Ricci scalar, $\mathcal{L_{M}}$ is the matter Lagrangian, $\phi$ describes a scalar field, $V(\phi)$ and $A(\phi)$, respectively, denote the self-interacting potential and conformal coupling to the matter field represented by $\Psi_{m}$. It must be noted that, in line with some of the $f(R)$ theories \cite{Babichev} and some low-energy string theories \cite{Bettoni}, the chameleon theory is also a scalar-tensor theory. However, due to the presence of non-canonical kinetic terms, the screening mechanisms such as k-mouflage \cite{Babichev1,Brax}, and Vainshtein \cite{Vainshtein,Burrage} can not be accommodated in Eq.~(\ref{eq1}). In the present framework, the scalar-matter interaction is mediated by the Jordan frame metric, $\tilde{g}_{ij}\equiv\,A^{2}(\phi)g_{ij}$. It is worth mentioning that the gravitational interaction gives rise to an effective fifth force between the matter fields, thereby placing the scalar-tensor framework under the scope of modified gravity \cite{Sakstein}. Now, to obtain the modified field equations in chameleon theory, we vary Eq.~(\ref{eq1}) with respect to the fundamental metric $g_{ij}$ and obtain:
\begin{equation}
    G_{ij}=8\pi\Bigg[T_{ij}+\nabla_{i}\phi\nabla_{j}\phi-g_{ij}\Bigg(\frac{1}{2}\nabla_{\chi}\phi\nabla^{\chi}\phi+V(\phi)\Bigg)\Bigg], \label{eq2}
\end{equation}
where the Einstein tensor is represented by $G_{ij}$ and, by definition of the matter Lagrangian, the energy-momentum tensor is expressed as: 
\begin{equation}
    T_{ij}=-\frac{2}{\sqrt{-g}}\frac{\mathcal{\delta L_{M}}}{\delta g^{ij}}. \label{eq3}
\end{equation}
Considering an isotropic and perfect fluid distribution, we can further express the energy-momentum tensor in the following form:
\begin{equation}
    T_{ij}=(\rho+p)u_{i}u_{j}+pg_{ij}. \label{eq4}
\end{equation}
In the rest frame of the fluid, $u^{i}u_{i}=-1$ represents the normalisation of the fluid four-velocity, $\rho$ is the energy density, and $p$ is the isotropic pressure component. Next, we vary Eq.~(\ref{eq1}) with respect to the scalar field $(\phi)$, and making use of the Jordan frame transformation, we obtain:
\begin{equation}
    \Box\,\phi=\frac{dV(\phi)}{d\phi}-T\frac{d\ln A(\phi)}{d\phi}, \label{eq5}
\end{equation}
where $T$ represents the trace of the energy-momentum tensor. Using the Bianchi identity, $\nabla^{j}G_{ij}=0$, we obtain the divergence of energy-momentum tensor in the following form:
\begin{equation}
    \nabla^{j}T_{ij}=\frac{d\ln A(\phi)}{d\phi}T\,\nabla_{i}\phi. \label{eq6}
\end{equation}
From Eq.~(\ref{eq6}), we note that in the Einstein frame, the energy-momentum tensor is not conserved, and hence the free particles do not follow the $g_{ij}$ geodesics. This may be attributed to Eq.~(\ref{eq5}), where the second term on the right-hand side produces the fifth force. However, we can achieve conformity with the Einstein frame by using the Jordan frame transformation. In the Jordan frame, the energy-momentum tensor can be defined in the following form:
\begin{equation}
    T_{ij}=-\frac{2}{\sqrt{-g}}\frac{\mathcal{\delta L_{M}}}{\delta \tilde{g}^{ij}}. \label{eq7}
\end{equation}
From Eqs.~(\ref{eq3}) and (\ref{eq7}), we note that $T_{ij}=A^{2}(\phi)\tilde{T}_{ij}$, and the conservation equation leads to $\nabla^{j}\tilde{T}_{ij}=0$. Hence, it can be argued that within the Jordan framework, the impact of the fifth force is embedded in the metric formalism. Moreover, these conformal transformations emerge from the tensor relations, as well as the four-velocity normalisation. In the Jordan frame, the latter becomes $u^{i}=A(\phi)\tilde{u}^{i}$ and subsequently the energy density and pressure are expressed in the following forms:
\begin{equation}
    \rho=A^{4}(\phi)\tilde{\rho},~~~~~~\text{and}~~~~~~p=A^{4}(\phi)\tilde{p}. \label{eq8}
\end{equation}
To ensure a relativistic description of the stellar configurations, we now consider a static, spherically symmetric space-time described by the following line element:
\begin{equation}
    ds^{2}=-e^{2\nu(r)}dt^{2}+e^{2\lambda(r)}dr^{2}+r^{2}d\theta^{2}+r^{2}sin^{2}\theta\,d\phi^{2}. \label{eq9}
\end{equation}
In the asymptotic limit, the space-time must approach a Schwarzschild one, and following Ref.~\cite{deAguiar}, we note that $e^{-2\lambda}=1-\frac{2m(r)}{r}$. Considering these ingredients and employing Eqs.~(\ref{eq2}), (\ref{eq5}) and (\ref{eq6}), we obtain the following set of equations:  
\begin{equation}
    \frac{d\nu}{dr}=\frac{1}{r-2m}\Bigg\{\frac{m}{r}+4\pi\,r^{2}\Bigg[A^{4}\tilde{p}+\Bigg(\frac{1}{2}-\frac{m}{r}\Bigg)\phi'^{2}-V\Bigg]\Bigg\}, \label{eq10}
\end{equation}
\begin{equation}
    \frac{dm}{dr}=4\pi\,r^{2}\Bigg[A^{4}\tilde{\rho}+\Bigg(\frac{1}{2}-\frac{m}{r}\Bigg)\phi'^{2}+V\Bigg], \label{eq11}
\end{equation}
\begin{equation}
    \frac{d\tilde{p}}{dr}=-(\tilde{p}+\tilde{\rho})\Bigg[\nu'+\frac{A_{,\phi}}{A}\phi'\Bigg], \label{eq12}
\end{equation}
and 
\begin{equation}
    \phi''=\Bigg[\frac{m'+\frac{3m}{r}-2}{r-2m}-\nu'\Bigg]\phi'+\frac{r}{r-2m}\Bigg[V_{,\phi}-A_{,\phi}A^{3}(3\tilde{p}-\tilde{\rho})\Bigg]. \label{eq13}
\end{equation}
Eqs.~(\ref{eq10})-(\ref{eq13}) are termed the modified TOV equations \cite{P6,P7}. Here, the overhead prime represents the derivatives with respect to $r$. For $A(\phi)=1$ and considering Eq.~(\ref{eq10}) up to $\mathcal{O}(2)$, one can retain the conventional TOV equations of GR. In the chameleon screening, the medium-dependent chameleon mass emerges from the $V(\phi)-A(\phi)$ interaction. Notably, the potential must have a monotonically decreasing nature from the centre to the surface such that it does not contain a localised minimum. However, due to its matter coupling, the combined effect may produce a minimum of the effective potential term on the right-hand side of Eq.~(\ref{eq5}). Following these arguments, we consider the following forms of the self-interacting potential $V(\phi)$ and conformal coupling parameter $(A(\phi))$ \cite{P8}:
\begin{equation}
    V(\phi)=\Lambda^{4}\Bigg(\frac{\Lambda}{\phi}\Bigg)^{n},~~~~~~\text{and}~~~~~~A(\phi)=e^{\sqrt{8\pi}\beta\phi}, \label{eq14}
\end{equation}
where $\Lambda$ is measured in the units of mass, $n>0$ and $\beta$ are constants. Moreover, $\beta$ characterises the magnitude of scalar-matter field interaction. On the other hand, $\Lambda$ is termed the chameleon energy scale \cite{P8} as it acts as a controlling parameter for the contribution of the scalar field to the universe's energy density. Notably, for $n,\,\beta=\mathcal{O}(1)$, the equivalence principle constrains $\Lambda$ in the range, $\Lambda\lesssim\frac{10^{-30}}{\sqrt{8\pi}}\approx1\,\text{millielectronvolt}$ \cite{P5}. This limit is exceptionally close to the characteristic dark energy scale responsible for the present cosmic acceleration.

\section{Realistic Equations of State and Mass-Radius Relations} \label{sec3}

\begin{table}[t!]
    \centering
    \caption{\raggedright Nuclear matter properties, {\it viz.} (i) Saturation number $(n_{s})$, (ii) Binding energy/nucleon $(E_{0})$, (iii) Incompressibility of symmetric NM $(K)$, (iv) Symmetry energy $(J)$, (v) Slope of the symmetry energy $(L)$, and (vi) Curvature of the symmetry energy $(K_{\mathrm{sym}})$, for BSk24 and SLy4 EoSs, respectively.}
    \label{tab1a}
    \begin{tabular}{ccccccc}
    \hline
    \hline
       \multirow{2}{*}{EoS} & $n_{s}$ & $E_{0}$  & $K$& $J$ & $L$ & $K_{\mathrm{sym}}$ \\
       & $(\mathrm{fm^{-3}})$ & (MeV) & (MeV) & (MeV) & (MeV) & (MeV) \\
        \hline
        BSk24 & 0.1578 & 16.048 & 245.5 & 30.0 & 46.4 & -37.6 \\
        \hline
        SLy4 & 0.159 & 15.97 & 230.0 & 32.04 & 46.0 & -119.73 \\
        \hline
        \hline
    \end{tabular}
\end{table}

To solve the system of Eqs.~(\ref{eq10})-(\ref{eq13}) with the model functions specified in Eq.~(\ref{eq14}), and to determine the corresponding maximum mass and radius (M-R), an appropriate EoS is required. A wide range of phenomenological and parametrised EoSs has been developed in the literature. Since our primary objective is to investigate the impact of the chameleon mechanism on neutron star structure, we employ the following realistic, state-of-the-art EoSs: 
\begin{itemize}
    \item PCP(BSk24): The Brussels-Montreal energy-density functional represented by the BSk24 EoS corresponds to a unified approach describing the cold, non-accreting neutron stars in beta equilibrium. BSk24 yields a maximum mass of $2.28~\mathrm{M_{\odot}}$ having an associated radius of $11.08$ Km \cite{Wang,Allard,Pearson,Pearson1,Pearson2,Goriely,Perot,Xu,Welker}.

Since astrophysical applications require extrapolating nuclear matter properties from experimentally constrained regimes to highly neutron-rich conditions, the employed energy-density functionals should incorporate the most reliable theoretical information available for neutron-rich matter. A key requirement is that the nuclear mass models remain consistent, up to densities characteristic of neutron star interiors, with the EoS of homogeneous pure neutron matter (NM) predicted by many-body calculations based on realistic two- and three-nucleon interactions. Although several EoSs have been developed, they exhibit substantial variations in their stiffness at suprasaturation densities. The motivation for employing the BSk24 EoS arises from the observation that simultaneously reproducing the NM EoS, nuclear masses, and other empirical properties of finite nuclei, such as charge radii, does not uniquely determine all the parameters of generalised Skyrme energy-density functionals. In particular, these constraints leave considerable freedom in specifying the nuclear symmetry energy, defined as the difference between the energy per nucleon of pure NM and that of symmetric nuclear matter (SNM):
\begin{equation}
    S(n)=e_{\mathrm{NM}}(n)-e_{\mathrm{SNM}}(n). \label{eq14a}
\end{equation}

This ambiguity primarily originates from the fact that nuclear mass data alone do not uniquely constrain the energy per nucleon of SNM, thereby permitting variations in the symmetry energy while maintaining agreement with existing empirical constraints. To characterise this freedom, the symmetry-energy coefficient $J$ is introduced through an expansion of the energy per nucleon of infinite nuclear matter (INM) around the saturation density $n_{s}$ and zero isospin asymmetry. Defining the dimensionless density deviation as $\epsilon=(n-n_{s})/n_{s}$ and the neutron–proton asymmetry parameter as $\eta=(n_{n}-n_{p})/n$, where $n_n$ and $n_p$ denote the neutron and proton number densities, respectively. Additionally, the expansion takes the following form:
\begin{equation}
    e(n,\eta)=E_{0}+
\left(J+\frac{1}{3}L\epsilon\right)\eta^2
+\frac{1}{18}\left(K+K_{\mathrm{sym}}\eta^2\right)\epsilon^2
+\cdots. \label{eq14b}
\end{equation}

Here, $E_{0}$ is the energy per nucleon at saturation, $J$ is the symmetry energy at saturation, $L$ denotes its density slope, $K$ is the incompressibility coefficient of symmetric nuclear matter, and $K_{\mathrm{sym}}$ characterises the curvature of the symmetry energy at saturation.

\item RG(SLY4): The SLy4 EoS likewise provides a unified description of matter in beta equilibrium at zero temperature. The maximum mass obtained in the model is $2.06~\mathrm{M_{\odot}}$ with a corresponding radius of 10.02 Km \cite{Gulminelli,Chabanat,Danielewicz}. The SLy4 EoS is based on the Skyrme effective nucleon-nucleon interaction, which was developed to describe the properties of finite nuclei and neutron-rich nuclear matter. In this framework, the effective interaction is expressed as
\begin{align}
V_{12}=&t_0(1+x_0P_\sigma)\delta(\mathbf{r})+\frac{t_1}{2}(1+x_1P_\sigma)
\left[\mathbf{k}'^2\delta(\mathbf{r})
+\delta(\mathbf{r})\mathbf{k}^2\right]+t_2(1+x_2P_\sigma)
\mathbf{k}'\cdot\delta(\mathbf{r})\mathbf{k}+\frac{t_3}{6}(1+x_3P_\sigma)
\rho^\alpha(\mathbf{R})\delta(\mathbf{r})
\nonumber\\&\quad+iW_0(\boldsymbol{\sigma}_1+\boldsymbol{\sigma}_2)
\cdot\left[\mathbf{k}'\times
\delta(\mathbf{r})\mathbf{k}\right],
\end{align}
where $\mathbf{r}=\mathbf{r}_1-\mathbf{r}_2$ and $\mathbf{R}=(\mathbf{r}_1+\mathbf{r}_2)/2$ denote the relative and centre-of-mass coordinates, respectively. Here, $P_\sigma$ is the spin-exchange operator, $\rho$ represents the nucleon number density, and $\mathbf{k}$ and $\mathbf{k}'$ are the relative-momentum operators acting to the right and left, respectively. The parameters $t_i$, $x_i$, $W_0$, and $\alpha$ specify the strength and density dependence of the interaction \cite{Chabanat}. These parameters define the Skyrme interaction and the associated energy-density functional, from which the thermodynamic properties of nuclear matter can be calculated. For applications to cold neutron stars, the matter composition is generally determined by imposing beta equilibrium and charge neutrality. The resulting energy density, $\varepsilon$, and pressure, $p$, establish the barotropic relation $p=p(\varepsilon)$ required for stellar-structure calculations. This EoS can subsequently be incorporated into the Tolman-Oppenheimer-Volkoff equations to obtain neutron star mass-radius sequences and investigate the corresponding macroscopic properties. The SLy4 parametrisation is therefore a useful framework for studying neutron stars within a microscopic nuclear-matter description.

For both the BSk24 and SLy4 EoSs, the properties of nuclear matter near saturation are tabulated in Table~\ref{tab1a}. 
\end{itemize} 

\begin{figure}[h!]
	\begin{minipage}{0.45\textwidth}
		\centering
		\includegraphics[width=8cm]{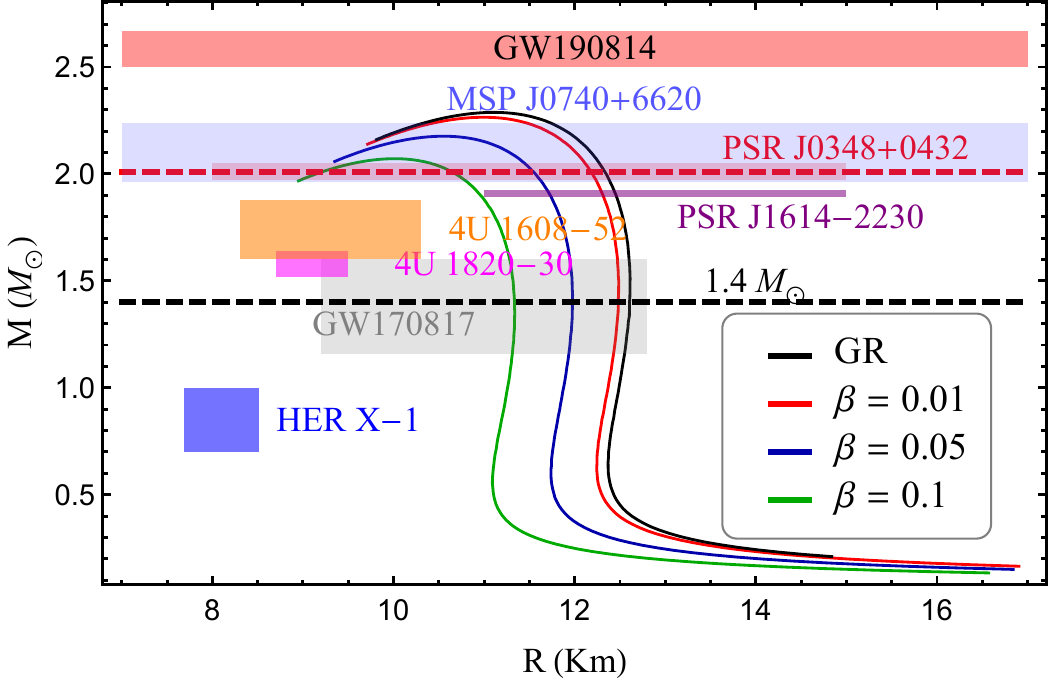}
		\subcaption{M-R plot using BSk24 EoS and different values of $\beta$.}
		\label{fig1}
	\end{minipage}
	\hfill
	\begin{minipage}{0.45\textwidth}
		\centering
		\includegraphics[width=8cm]{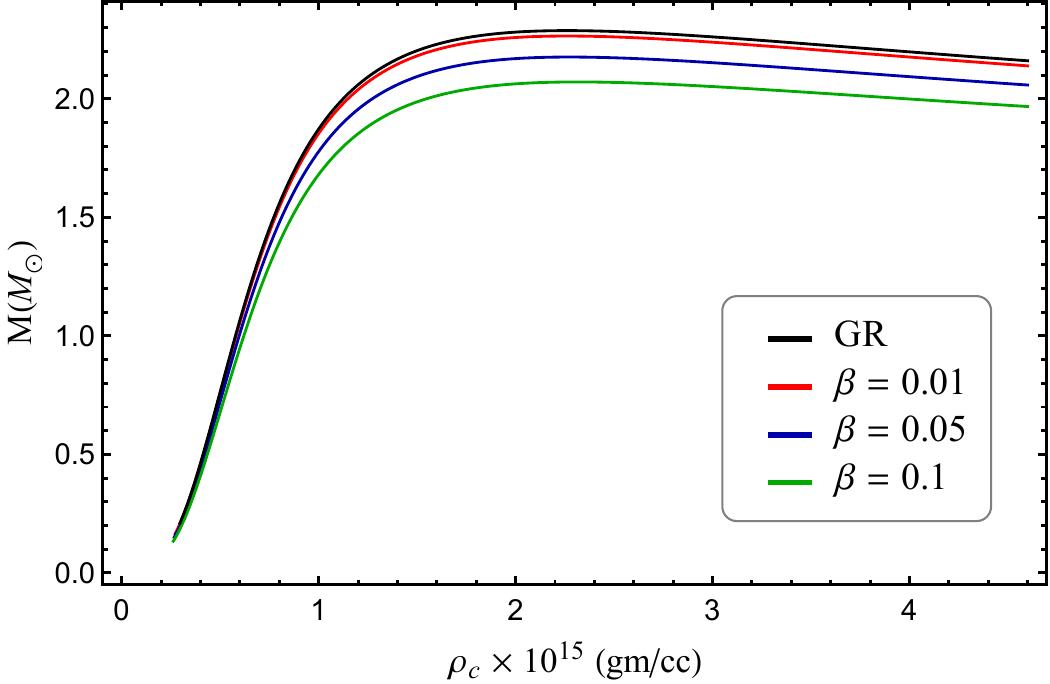}
		\subcaption{\raggedright Mass vs. central density plot using BSk24 EoS and different values of $\beta$.}
		\label{fig2}
	\end{minipage}
	\centering
	\caption{\raggedright M-R and mass-central density relations obtained from the TOV equations in Chameleon theory for the BSk24 EoS. The GR limit yields a maximum mass of $2.28\,\mathrm{M_{\odot}}$ and a corresponding radius of $11.09\,\mathrm{Km}$. Here, $\Lambda=10^{-18}M_p$ and $n=1$, with $M_p$ being the Planck mass.}
	\label{fig1a}
\end{figure}

\begin{figure}[h!]
	\begin{minipage}{0.45\textwidth}
		\centering
		\includegraphics[width=8cm]{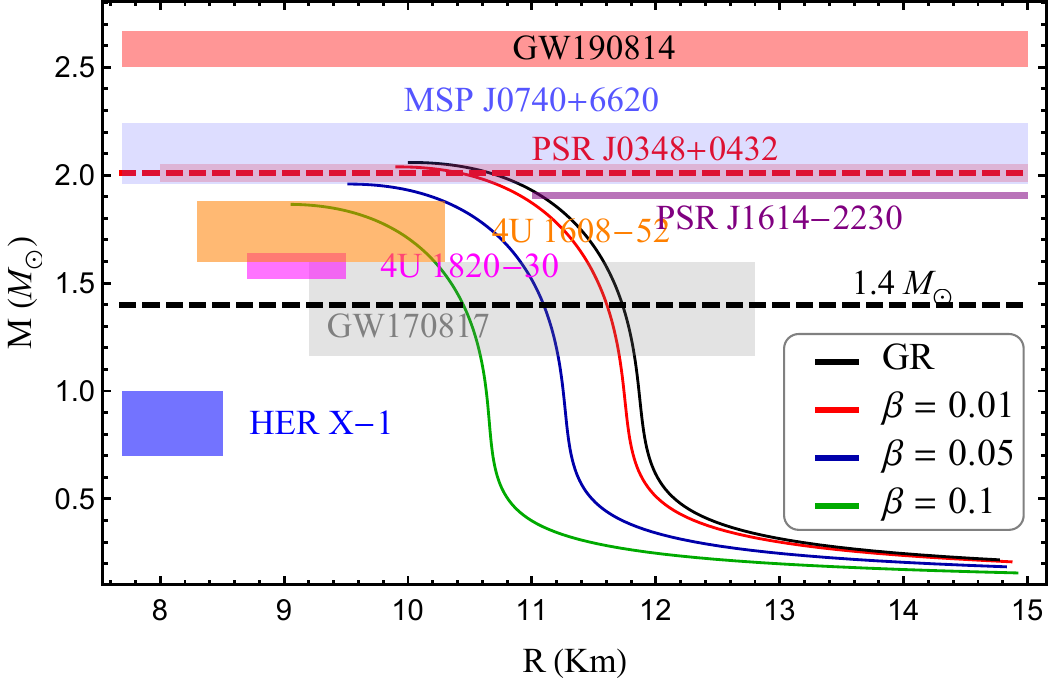}
		\subcaption{\raggedright M-R plot using SLy4 EoS and different values of $\beta$.}
		\label{fig3}
	\end{minipage}
	\hfill
	\begin{minipage}{0.45\textwidth}
		\centering
		\includegraphics[width=8cm]{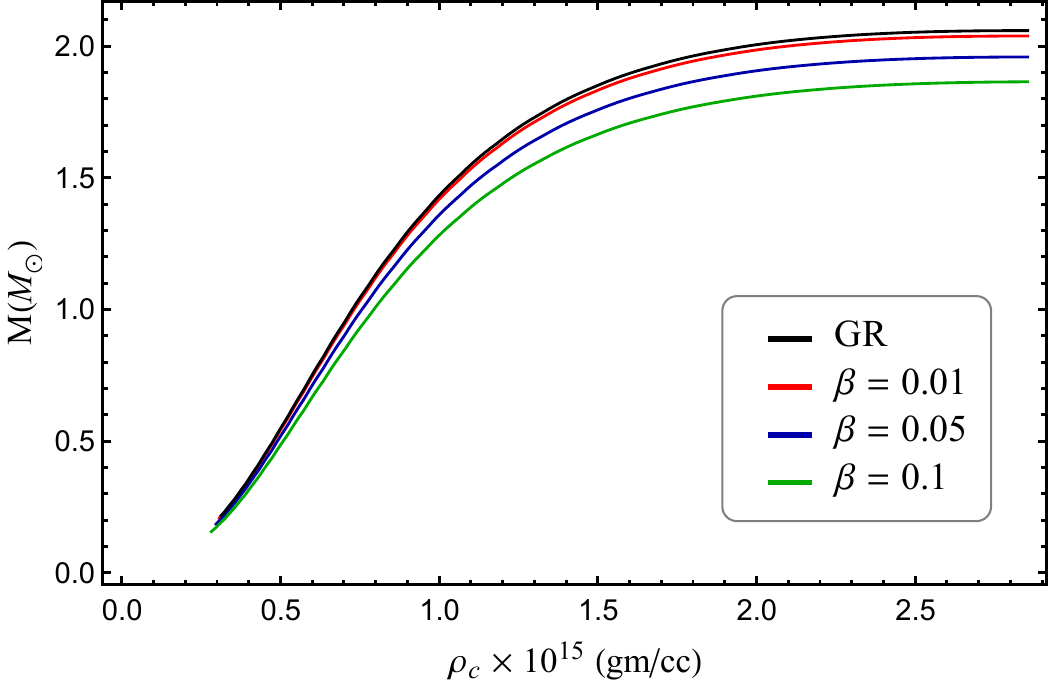}
		\subcaption{\raggedright Mass vs. central density plot using SLy4 EoS and different values of $\beta$.}
		\label{fig4}
	\end{minipage}
	\centering
	\caption{\raggedright M-R and mass-central density relations obtained from the TOV equations in Chameleon theory for the SLy4 EoS. The GR limit yields a maximum mass of $2.06\,\mathrm{M_{\odot}}$ and a corresponding radius of $10\,\mathrm{Km}$. Here, $\Lambda=10^{-18}M_p$ and $n=1$.}
	\label{fig2a}
\end{figure}

Apart from the EoSs, we impose the following boundary conditions, $m(r=0)=0,\,\tilde{p}(r=R)=0,\,\text{and}\,\tilde{\rho}(r=0)=\text{central density}$, and the obtained results are illustrated in Figs.\ref{fig1a} and \ref{fig2a}.


By varying the scalar-matter field interaction strength through $\beta$, we have shown the M-R relationship in Figs.~\ref{fig1} and \ref{fig3}, while the mass-central density profiles are demonstrated in Figs.~\ref{fig2} and \ref{fig4}. Moreover, from Figs.~\ref{fig1} and \ref{fig3}, we have extracted the maximum M-R values for different choices of $\beta$, and the results are tabulated in Tables~\ref{tab1} and \ref{tab2}. 


\begin{table}[h!]
\caption{\raggedright Tabulation of maximum mass and radius for (a) BSk24 EoS and (b) SLy4 EoS considering different values of $\beta$. For lower values of $\beta$, we retain the maximum mass bounds for the respective EoSs.}
    \begin{minipage}{0.45\textwidth}
    \centering
	\label{tab1}
	\begin{tabular}{ccc}
		\hline
        \hline
		$\beta$ & M $\mathrm{(M_{\odot})}$ & R (Km) \\
		\hline
		0.01 & 2.26 & 10.98 \\
		0.05 & 2.17 & 10.56 \\
		0.1 & 2.07 & 10.02 \\
		\hline
        \hline
	\end{tabular}
    \subcaption{}
    \end{minipage}
\hfil
\begin{minipage}{0.45\textwidth}
	\centering
	\label{tab2}
	\begin{tabular}{ccc}
		\hline
        \hline
		$\beta$ & M $\mathrm{(M_{\odot})}$ & R (Km) \\
		\hline
		0.01 & 2.04 & 9.91 \\
		0.05 & 1.96 & 9.52 \\
		0.1 & 1.86 & 9.06 \\
		\hline
        \hline
	\end{tabular}
    \subcaption{}
    \end{minipage}
\end{table}


It is noted that in both cases, the maximum mass decreases with increasing $\beta$. This may be attributed to the fact that with increasing $\beta$, the strength of scalar-matter field coupling increases. Consequently, the compressibility increases and the stellar structure approaches a more compact form. On the other hand, with increasing compressibility, the EoSs approach a softer nature. The combined effects of these two features may reduce the maximum mass and the associated radius in the model. Additionally, we have also observed that the obtained maximum mass limits are consistent with the recently observed pulsars and the low-mass companions of GW events. Hence, we have incorporated the following benchmark results in our M-R plots:

\begin{itemize}
    \item \textit{Most massive pulsars:} (i) NANOGrav data \cite{Arzoumanian} collected from millisecond pulsars have determined the pulsar-binary component mass of PSR J1614-2230 to be   $1.908\pm0.016~\mathrm{M_{\odot}}$, (ii) the relativistic Shapiro delay measurements have revealed that MSP J0740+6620 may be termed the most massive neutron star observed, to date, having a mass of $2.14^{+0.10}_{-0.09}~\mathrm{M_{\odot}}$. \cite{Cromartie}, and (iii) using radio-timing observations, Antoniadis et al. \cite{Antoniadis} obtained a mass of $2.01\pm0.04~\mathrm{M_{\odot}}$ for PSR J0348+0432. 
    
    \item \textit{GW170817:} The first binary-neutron-star merger detected by the Advanced LIGO-Virgo network \cite{Abbott} had a total detector-frame mass of approximately $2.74^{+0.04}_{-0.01}~\mathrm{M_{\odot}}$, with the component masses constrained to lie in the range $1.17-1.60~\mathrm{M_{\odot}}$.

    \item \textit{Radius constraint:} Based on a combination of multi-messenger observations and nuclear-theory constraints, Capano et al. \cite{Capano} inferred a radius of $11.0^{+0.9}_{-0.6}~\mathrm{km}$ for a $1.4~\mathrm{M_{\odot}}$ neutron star.

    \item \textit{Additional massive compact objects:} We also indicate the relevant mass ranges associated with the secondary component of GW190814 \cite{Abbott1} ($2.59^{+0.08}_{-0.09}~\mathrm{M_{\odot}}$) and several observed X-ray pulsars, including 4U 1608-52 \cite{Guver}, 4U 1820-30 \cite{Guver1}, and Her X-1 \cite{Abubekerov}. In particular, the inferred mass of the GW190814 secondary lies above the maximum masses obtained for the present stellar configurations. 
\end{itemize}

It must be noted that in the chameleon mechanism, $\beta$ plays a central role in characterising a stellar structure. Hence, it is imperative to study its allowed parameter range. In this regard, we use the constraints from the most massive pulsars and multi-messenger observations, as demonstrated in Figs.~\ref{fig1} and \ref{fig3}. Consequently, we obtain the following:


\begin{table}[h!]
	\centering
	\caption{Constraining the range of $\beta$ through massive pulsars.}
	\label{tab3}
	\begin{tabular}{cccc}
		\hline
        \hline
	\multirow{2}{*}{EoSs} & \multirow{2}{*}{Pulsars} & Observed mass & \multirow{2}{*}{Range of $\beta$} \\
    && $(\mathrm{M_{\odot}})$ & \\
		\hline
	\multirow{3}{*}{BSk24} & PSR J1614-2230 & 1.924-1.892 & 0.176-0.1936 \\
	&	PSR J0348+0432 & 2.05-1.97 & 0.11-0.15 \\
	&	MSP J0740+6620 & 2.24-2.05 & 0.021-0.11 \\
	\hline
	\multirow{3}{*}{SLy4} & PSR J1614-2230 & 1.924-1.892 & 0.068-0.085\\
	&	PSR J0348+0432 & 2.05-1.97 & 0.004-0.044\\
	&	MSP J0740+6620 & 2.24-2.05 & -- \\
	\hline
    \hline
	\end{tabular}
\end{table}

\begin{figure}[h!]
	\begin{minipage}{0.45\textwidth}
		\centering
		\includegraphics[width=8cm]{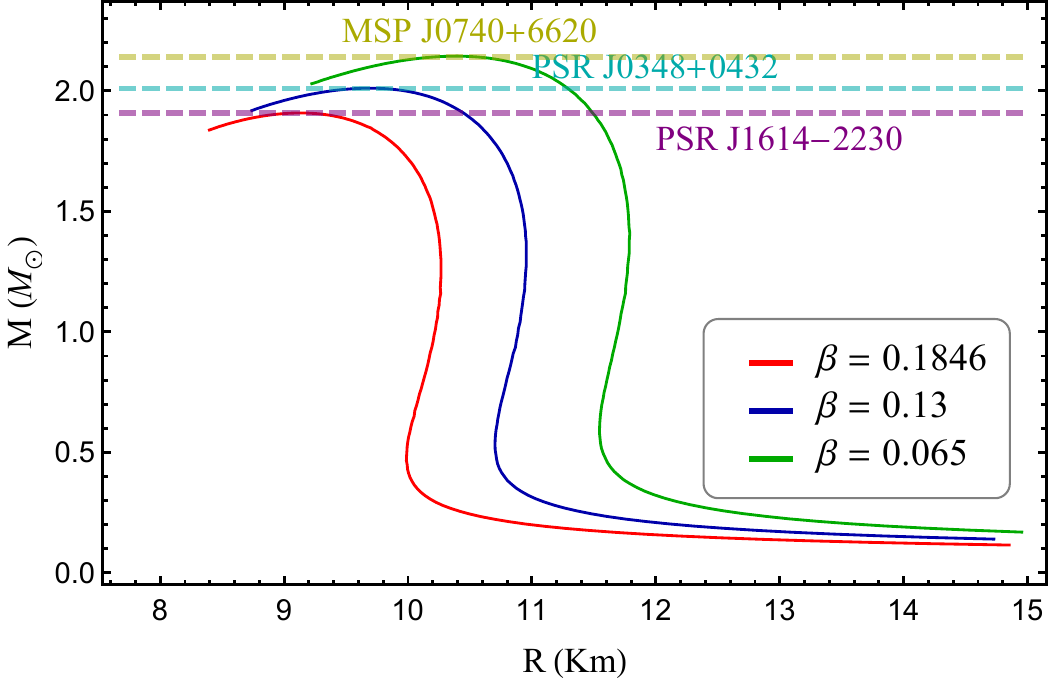}
		\subcaption{}
		\label{fig5}
	\end{minipage}
	\hfill
	\begin{minipage}{0.45\textwidth}
		\centering
		\includegraphics[width=8cm]{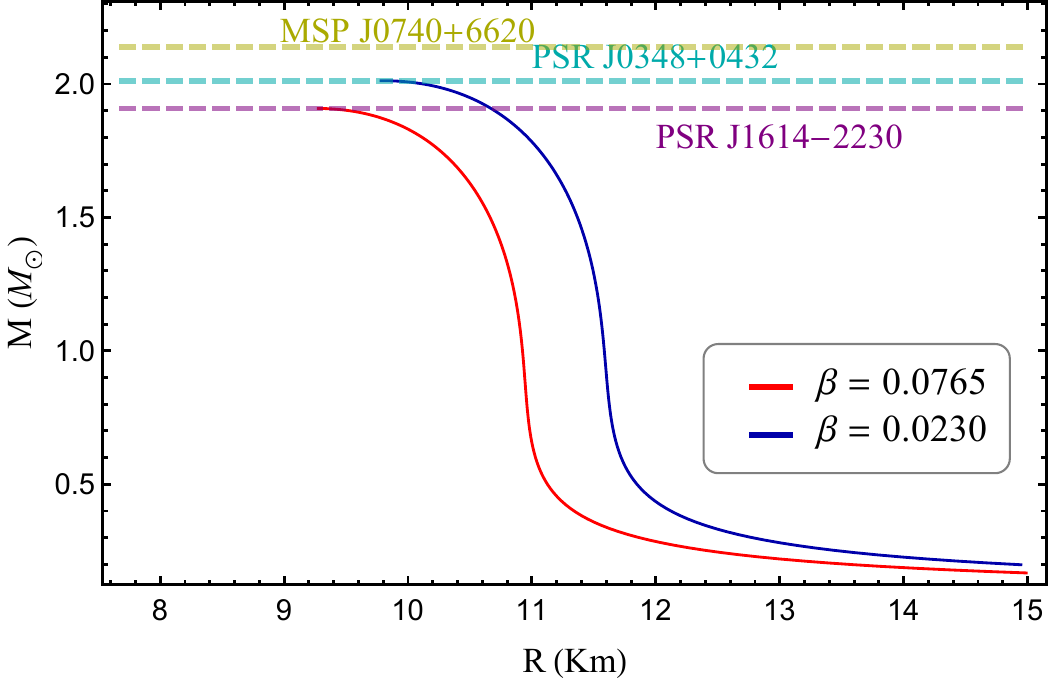}
		\subcaption{}
		\label{fig6}
	\end{minipage}
	\centering
	\caption{\raggedright M-R plots and the mass ranges of massive pulsars for (a) BSk24 EoS, and (b) SLy4 EoS with different values of $\beta$. Here, $\Lambda=10^{-18}M_p$ and $n=1$.}
	\label{fig3a}
\end{figure}

With regard to the massive pulsar-mass measurements, Table~\ref{tab3} shows that the permissible range of $\beta$ for the BSk24 EoS is 0.021-0.1936 and for the SLy4 EoS is 0.004-0.085. Since these neutron stars are among the most massive precisely measured compact objects to date, the corresponding ranges of $\beta$ may be regarded as physically viable parameter ranges within the chameleon theory. Within the present parameter space, we fine-tuned $\beta$ to obtain the maximum masses of massive pulsars for both EoSs by solving the TOV equations. The corresponding results are shown in Fig.~\ref{fig3a}. 


Notably, the maximum mass in the case of SLy4 EoS is capped at $2.06~\mathrm{M_{\odot}}$. Hence, an explanation of MSP J0740+6620 is not possible using SLy4 EoS. Further, the study of $1.4~\mathrm{M_{\odot}}$ neutron stars \cite{Capano} yields the following bounds: 


\begin{table}[h!]
	\centering
	\caption{Range of $\beta$ to predict the radius $R_{1.4~\mathrm{M_{\odot}}}$=11.9-10.4 Km \cite{Capano}.}
	\label{tab4}
	\begin{tabular}{cc}
		\hline
        \hline
		EoS & $\beta$\\\hline
		BSk24 & 0.056-0.173\\
		SLy4 & 0.00001 (11.72 Km)-0.103\\
		\hline
        \hline
	\end{tabular}
\end{table}


From Table~\ref{tab4}, we note two interesting features: (i) for BSk24, the obtained range of $\beta$ lies within the limits obtained in Table~\ref{tab3}. This feature further reinforces the viability of the obtained bounds, (ii) in the case of SLy4 EoS, we can achieve a radius as high as 11.72 Km for $\beta$ as low as $0.00001$. Hence, the range can be further refined to 0.00001-0.085, with a caveat that SLy4 does not produce the upper limit of $R_{1.4~\mathrm{M_{\odot}}}$. However, we can accurately predict the median radius of $R_{1.4~\mathrm{M_{\odot}}}$ by tuning $\beta$, as shown in Table~\ref{tab5}. 


\begin{table}[h!]
	\centering
	\caption{\raggedright Predicted radius of $1.4\,\mathrm{M_{\odot}}$ neutron star from BSk24 and SLy4 EoSs. The median radius of $11.0^{+0.9}_{-0.6}$ Km has been considered as the reference. \cite{Capano}}
	\label{tab5}
	\begin{tabular}{ccc}
		\hline
        \hline
		EoS & $\beta$ & Predicted radius for $R_{1.4\,\mathrm{M_{\odot}}}$ (Km)\\
		\hline
		BSk24 & 0.126 & 11.00 \\
		SLy4 & 0.057 & 11.00 \\
		\hline
        \hline
	\end{tabular}
\end{table}


\section{Numerical results}

The results presented in Figs. 1–3 and Tables I–IV demonstrate that the chameleon mechanism can have a significant impact on the macroscopic properties of neutron stars, even when realistic equations of state are employed. In particular, the scalar–matter coupling parameter $\beta$ controls the departure from the corresponding general-relativistic configurations, affecting both the maximum mass and the stellar radius. The comparison between the BSk24 and SLy4 equations of state further illustrates the importance of accounting for nuclear-matter uncertainties when constraining modified theories of gravity through compact-star observations.

\smallskip

Figures 1(a) and 2(a) show that the mass–radius relations shift towards lower masses and smaller radii as $\beta$ increases. This behaviour is quantified in Table I. For the BSk24 EoS, the maximum mass decreases from $2.28~M_\odot$ in the general-relativistic limit to $2.26~M_\odot$, $2.17~M_\odot$, and $2.07,M~\odot$ for $\beta=0.01, 0.05$, and 0.1, respectively. The corresponding radius decreases from 11.09 km to 10.98, 10.56, and 10.02 km. A similar trend is observed for SLy4, for which the maximum mass decreases from $2.06,M~\odot$ in GR to $2.04~M_\odot$, $1.96~M_\odot$, and $1.86~M_\odot$, while the corresponding radius decreases from $10.00~km$ to 9.91, 9.52, and $9.06~km$.

\smallskip

These results indicate that the stellar response to the chameleon coupling is systematic for both nuclear EoSs. At $\beta=0.1$, for example, the maximum mass is reduced by approximately $9.2\%$ relative to its general-relativistic value for BSk24 and by approximately $9.7\%$ for SLy4. The corresponding reductions in radius are approximately $9.6\%$ and $9.4\%$, respectively. Thus, although the absolute masses and radii depend on the adopted EoS, the relative changes induced by the coupling are comparable in the two cases. This finding suggests that the qualitative influence of the chameleon field is not restricted to a particular nuclear-matter prescription.

\smallskip

The mass–central-density relations displayed in Figs. 1(b) and 2(b) provide a complementary view of this behaviour. Increasing $\beta$ lowers the maximum mass attained along the equilibrium sequences, indicating that the coupling modifies the range of central densities over which configurations with increasing mass can be supported. These curves are useful for identifying the maximum-mass configurations, although a complete assessment of their stability would require an explicit stability analysis. In particular, the location of the maximum mass along an equilibrium sequence is commonly used as an indicator of the onset of radial instability, but the precise stability criterion in a scalar–tensor theory should be established consistently with the scalar-field dynamics and the relevant boundary conditions.

\smallskip

The comparison between the two EoSs is particularly instructive. In the general-relativistic limit, BSk24 supports a maximum mass of $2.28~M_\odot$, whereas SLy4 yields $2.06~M_\odot$. The latter value is already close to the lower end of the mass interval associated with PSR J0348+0432 and below the central mass estimate of MSP J0740+6620. The additional reduction caused by the chameleon coupling therefore has different observational consequences for the two EoSs. While BSk24 retains the capacity to accommodate the most massive precisely measured pulsars over part of the parameter space, SLy4 becomes increasingly constrained as $\beta$ grows. This comparison highlights an important degeneracy between the properties of dense nuclear matter and modifications of gravity: a change in the maximum mass cannot, by itself, be attributed uniquely to the gravitational theory without accounting for the underlying EoS.

\smallskip

The constraints reported in Table II quantify this dependence on the observational benchmarks. For BSk24, the individual pulsar mass intervals select different ranges of $\beta$, with the largest masses imposing the strongest restrictions on the coupling. In particular, the range inferred from MSP J0740+6620 extends to smaller values of $\beta$ than the range obtained from PSR J1614-2230. This is consistent with the decreasing maximum mass found in Table I: sufficiently large coupling values prevent the equilibrium sequence from reaching the mass required by the observation. For SLy4, the mass of PSR J1614-2230 can be accommodated only over a restricted coupling interval, while the mass of PSR J0348+0432 provides a separate constraint. Within the parameter space investigated here, the maximum mass of the SLy4 sequence does not reach the mass interval adopted for MSP J0740+6620. Consequently, this particular EoS–gravity combination cannot reproduce that benchmark through a choice of $\beta$ alone.

\smallskip

It is important to interpret these intervals as constraints obtained within the assumed chameleon model, rather than as model-independent bounds on modified gravity. They depend on the selected EoS, the adopted observational mass intervals, and the fixed values $\Lambda=10^{-18}M_{\rm P}$ and $n=1$. In addition, the individual intervals in Table II should not automatically be interpreted as a single jointly allowed interval: a simultaneous constraint must be obtained from the intersection of the ranges required by all observations that the model is intended to satisfy.

\smallskip

The radius constraint for a $1.4~M_\odot$ neutron star provides an independent test of the model. Table III considers the observational interval $R_{1.4}=(10.4\text{--}11.9)~km$, while Table IV identifies coupling values for which the predicted radius equals the reference value of 11.0 km. The reported values, $\beta=0.126$ for BSk24 and $\beta=0.057$ for SLy4, show that the same reference radius can be reproduced by different combinations of EoS and scalar–matter coupling. This is another manifestation of the degeneracy between nuclear physics and modified gravity. Nevertheless, reproducing a single radius value is not sufficient to establish that a parameter choice satisfies all observational constraints. The radius-selected values of $\beta$ must also be checked against the mass constraints, and the full predicted radius interval should be compared with the observational range rather than relying solely on its median.

\smallskip

Taken together, the results show that realistic neutron-star configurations can provide complementary constraints on chameleon gravity through their maximum masses and radii. The BSk24 EoS remains compatible with a broader set of massive-pulsar benchmarks in the parameter ranges considered, whereas SLy4 is more restrictive, particularly for the most massive observed pulsars. At the same time, the possibility of obtaining similar radii with different EoSs and coupling strengths demonstrates why mass and radius information should be considered jointly. A more complete constraint analysis could combine several mass and radius measurements in a consistent statistical framework, while varying the chameleon energy scale and potential exponent in addition to $\beta$. Such an analysis would help establish whether the apparent compatibility of particular parameter ranges persists when uncertainties in both nuclear matter and the chameleon sector are taken into account.

\section{Scalar field, normalisation and boundary conditions}\label{sec5}
Regularity of the solution at the stellar centre requires a vanishing scalar-field gradient, $\phi'(0)=0$, whereas the central value $\phi(0)$ is not initially fixed. What is known is the asymptotic behaviour, $\phi\to\phi_{\infty}$ and $\phi'\to 0$ as $r\to\infty$. We thus treat the problem as a boundary-value problem: the same TOV system is integrated outwards from the centre, matched to the vacuum exterior, and $\phi(0)$ is tuned with a secant iteration until the asymptotic value $\phi_{\infty}$ is recovered within a prescribed tolerance.

Because matter couples to the Jordan-frame metric
$\tilde g_{\mu\nu}=A^2(\phi)g_{\mu\nu}$, a non-zero $\phi_{\infty}$ changes the gravitational strength measured at infinity to $G_{\rm eff}=A^2(\phi_\infty)G$. For a uniform field, the structure equations are mapped onto their GR form by $r\to A^{-2}r$ and $m\to A^{-2}m$, so that masses and radii are rescaled by $A^{-2}(\phi_{\infty})$. Consequently, the bulk of the shift in the
maximum mass and the corresponding radius relative to GR reflects the choice of $\phi_{\infty}$ rather than a change in stellar structure. To isolate the effect of the field profile, the scalar-tensor and GR stars should be compared at the same locally measured $G$, i.e. with $M$ and
$R$ expressed in units rescaled by $A^2(\phi_{\infty})$. The residual deviation, which is small, is attributable to the scalar-field gradient inside the star.

\begin{figure}
	\centering 
	\begin{minipage}{0.45\textwidth}
		\centering
		\includegraphics[width=8cm]{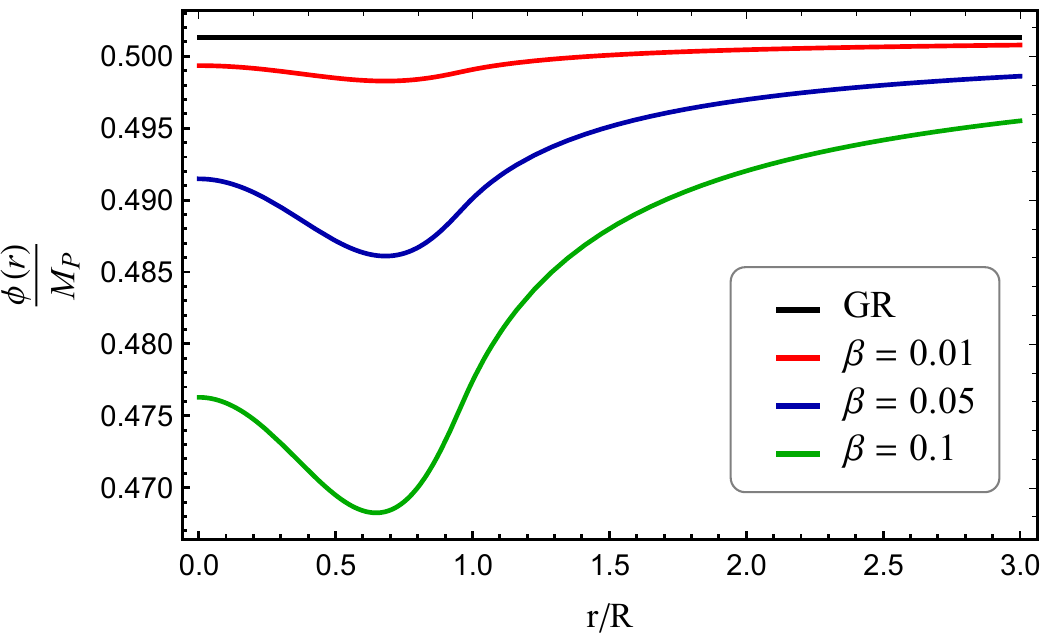}
	\end{minipage}
	\hfill
	\begin{minipage}{0.45\textwidth}
		\centering
		\includegraphics[width=8cm]{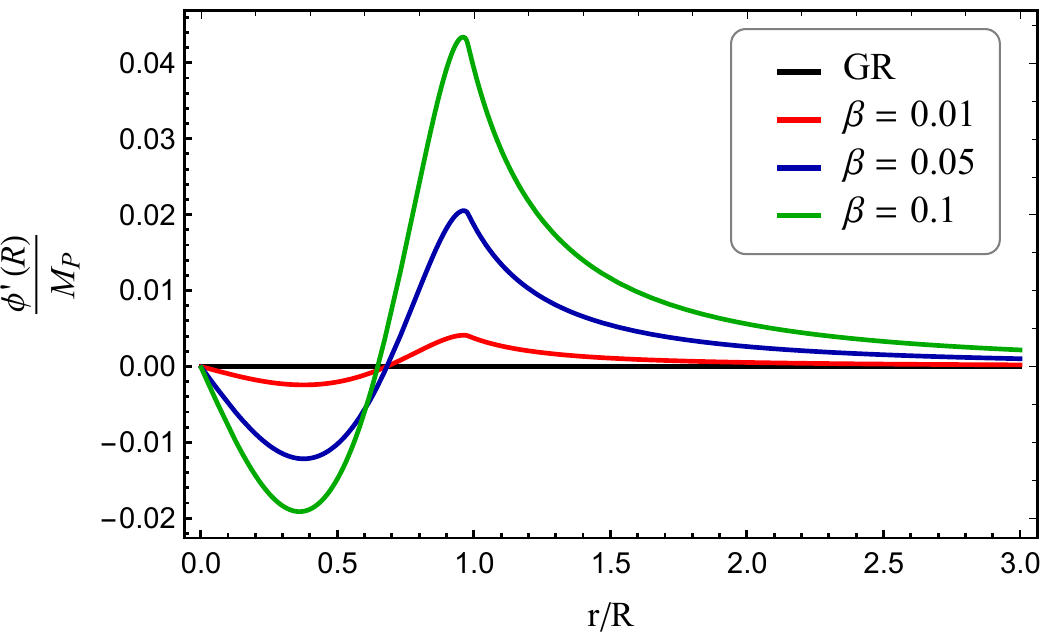}\\
	\end{minipage}
	\caption{Scalar field and field gradient profiles for BSk24 EoS. Here, $\Lambda=10^{-18}M_p$ and $n=1$.}
    \label{fig4aa}
\end{figure}

\begin{figure}
	\centering 
	\begin{minipage}{0.45\textwidth}
		\centering
		\includegraphics[width=8cm]{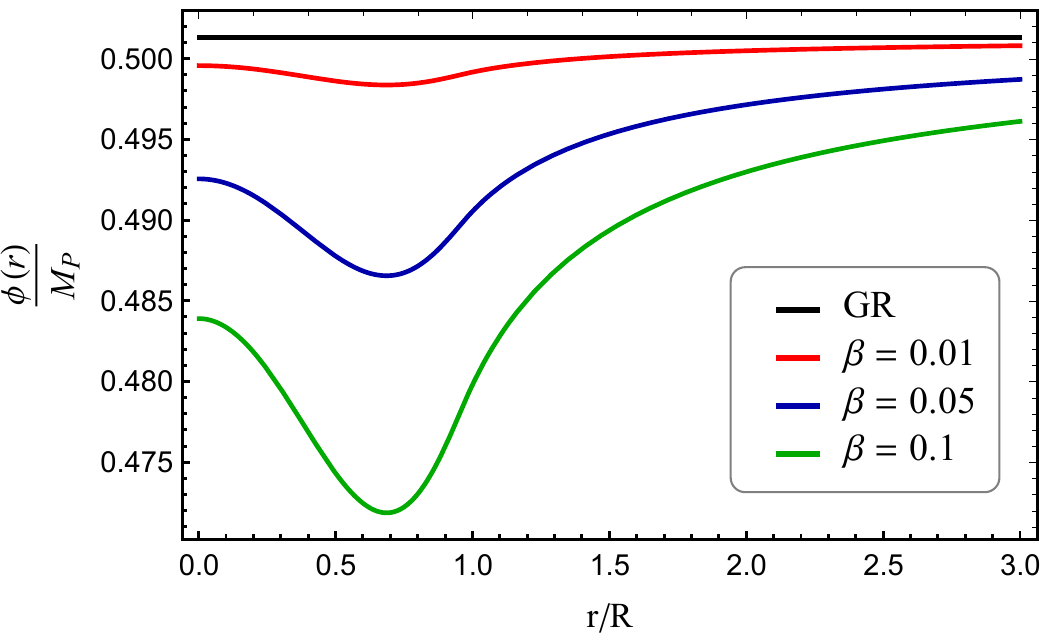}
	\end{minipage}
	\hfill
	\begin{minipage}{0.45\textwidth}
		\centering
		\includegraphics[width=8cm]{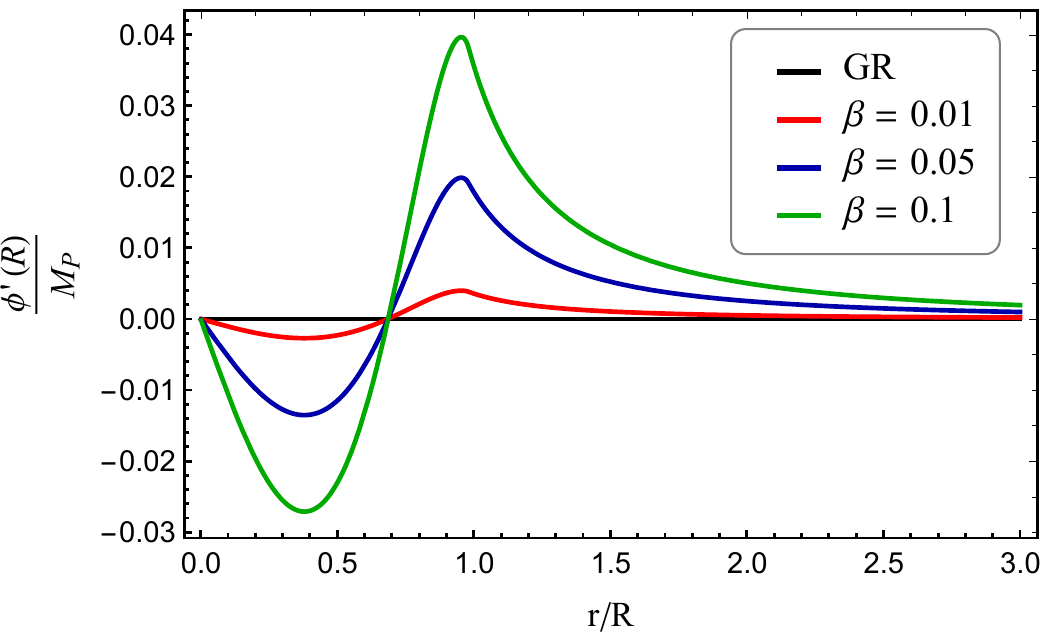}
	\end{minipage}
	\caption{Scalar field and field gradient profiles for SLy4 EoS. Here, $\Lambda=10^{-18}M_p$ and $n=1$.}
    \label{fig5aa}
\end{figure}

The radial profiles of the scalar field and its gradient are presented in Figs.~\ref{fig4aa} and \ref{fig5aa}: the field is expressed in units of the
reduced Planck mass, $\phi/M_{P}$, the radial coordinate is scaled by the stellar radius, $r/R$, and the gradient is shown as $\phi'(r)/M_{P}$. The results are illustrated for BSk24 and SLy4 EoSs for different values of the scalar-matter interaction parameter, $\beta$. 

Figs.~\ref{fig4aa} and \ref{fig5aa} show that, for both the EoSs, the scalar field
approaches its prescribed asymptotic value, $\phi\to\phi_{\infty}$, and its gradient decays to zero at large distance. For $\beta=0$ the field remains constant, as in GR, whereas for $\beta>0$ the amplitude of the gradient
grows with the coupling. The normalised gradient $\phi'(R)/M_{P}$ is negative in the inner region and becomes positive at larger radii. This behaviour follows from the source term of Eq.~(\ref{eq13}), $A_{,\phi}A^{3}(\tilde{\rho}-3\tilde{p})$, which is proportional to the trace of the matter energy-momentum tensor. In the dense core of a neutron star the pressure is comparable to the energy density, and hence $\tilde{\rho}-3\tilde{p}<0$. Moreover, $\phi'(r)$, which vanishes at the centre, is driven to negative values. Moving outward, the pressure falls below $\tilde{\rho}/3$, the source changes sign, and $\phi'$ crosses zero and increases to a maximum near $r/R\approx1$. In the vacuum exterior, the source vanishes and the
solution satisfies $r(r-2M)\,\phi'=\text{const}$, so that $\phi'\propto 1/[r(r-2M)]$. 

It must also be noted that the sign change affects the hydrostatic structure. In Eq.~(\ref{eq12}), the term $(A_{,\phi}/A)\,\phi'$ in $d\tilde{p}/dr$ reduces the pressure gradient where $\phi'<0$ and enhances it where $\phi'>0$. Since the field itself is non-monotonic, it decreases slightly in the core before increasing monotonically towards $\phi_{\infty}$, as seen in Figs.~\ref{fig4aa} and \ref{fig5aa}.

\section{Analysing stability through Harrison-Zel'dovich-Novikov criterion} \label{sec6}

\begin{figure}[h!]
	\begin{minipage}{0.45\textwidth}
		\centering
		\includegraphics[width=8cm]{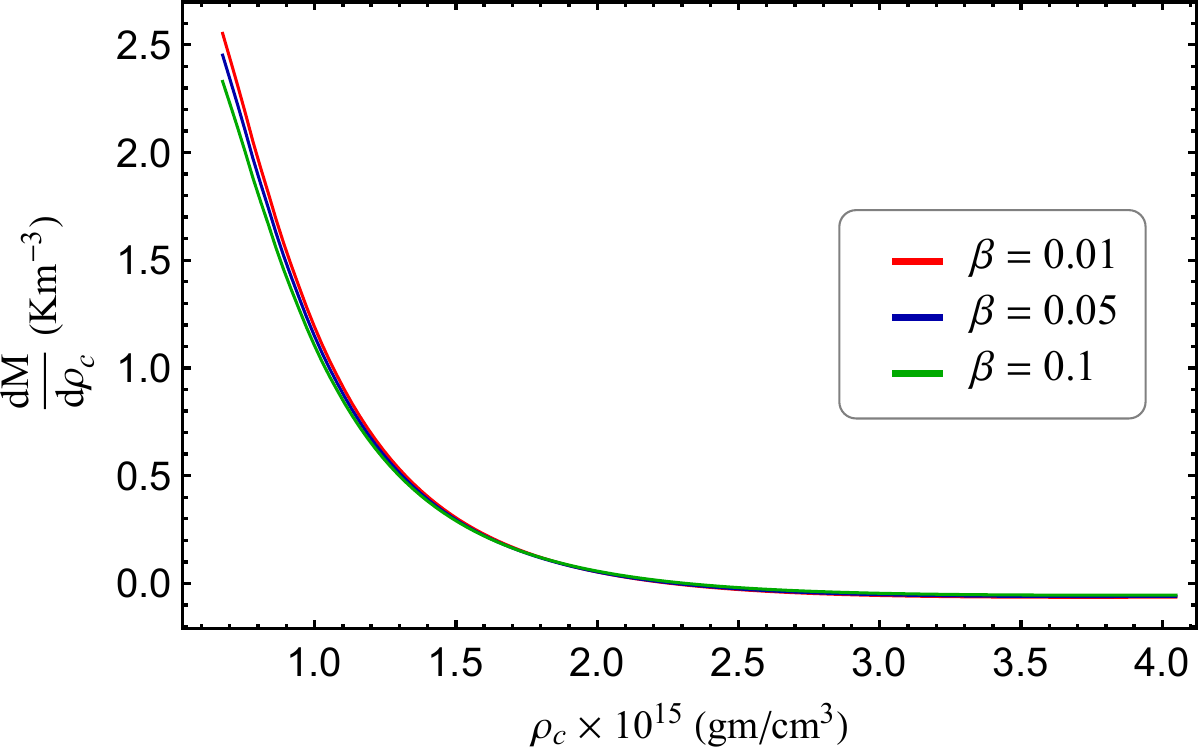}
		\subcaption{}
		\label{fig7}
	\end{minipage}
	\hfill
	\begin{minipage}{0.45\textwidth}
		\centering
		\includegraphics[width=8cm]{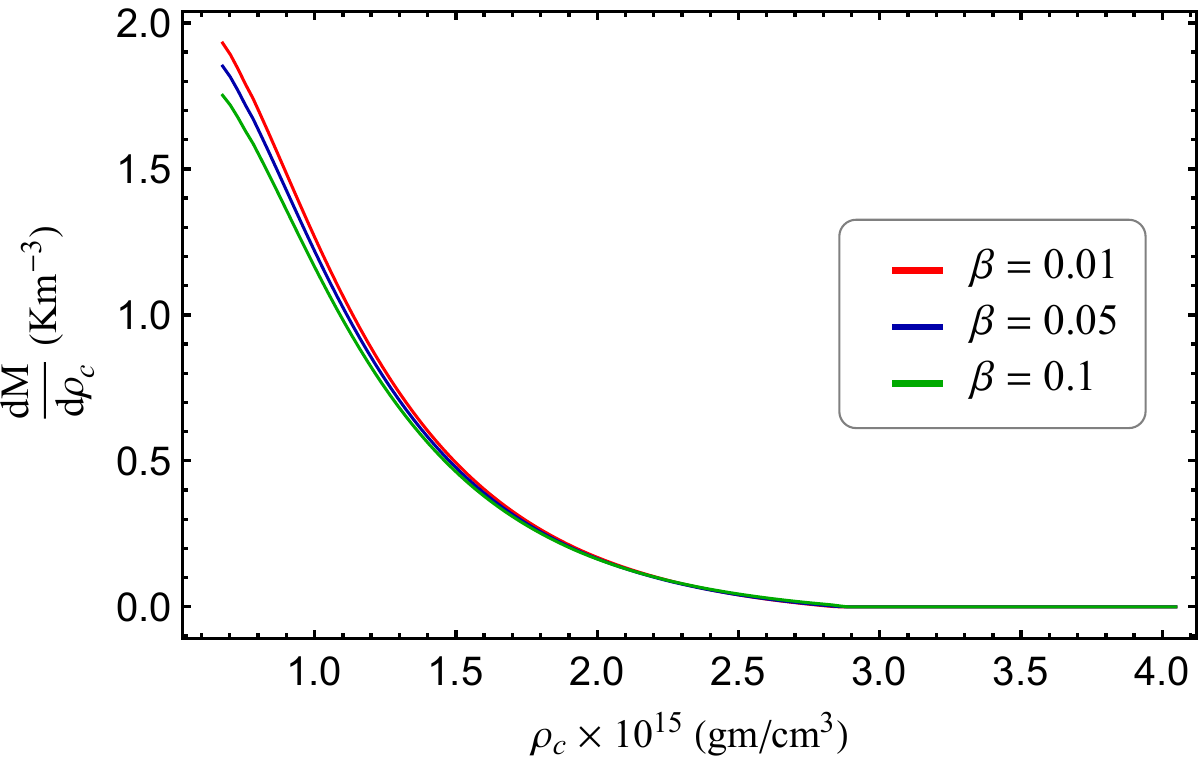}
		\subcaption{}
		\label{fig8}
	\end{minipage}
	\centering
	\caption{\raggedright $\frac{dM}{d\rho_{c}}$ vs. $\rho_{c}$ diagrams for (a) BSk24 EoS, and (b) SLy4 EoS with different values of $\beta$. Here, $\Lambda=10^{-18}M_p$ and $n=1$.}
	\label{fig4a}
\end{figure}

The Harrison-Zel'dovich-Novikov (HZN) criterion \cite{Harrison,Zeldovich} provides a widely used framework for assessing the stability of equilibrium stellar configurations against radial perturbations. It identifies the onset of instability along a sequence of stellar models and is closely associated with the behaviour of the fundamental radial oscillation mode (f-mode), whose squared frequency approaches zero at the stability boundary. In practice, this criterion is commonly investigated by examining the dependence of the gravitational mass, $M$, on the central density, $\rho_{c}$, particularly in the vicinity of the maximum-mass configuration. By integrating the TOV equations for successively increasing central densities, one can construct a sequence of equilibrium stellar models and identify the configuration at which the gravitational mass reaches its maximum. Beyond this turning point, the sequence generally enters an unstable branch, where the pressure support is no longer sufficient to maintain stability against radial perturbations, potentially leading to gravitational collapse. Since our preceding analysis demonstrates that the chameleon mechanism influences the internal structure and dynamics of neutron stars described by realistic EoSs, consequently modifying their mass-radius relationships, it is essential to examine how this modification affects the stability of the maximum-mass configurations across the parameter space considered in this study. The corresponding results are presented graphically in Figs.~\ref{fig7} and \ref{fig8}.

From Fig.~\ref{fig4a}, we infer that the maximum mass points described in Table~\ref{tab1} together with Figs.~\ref{fig2} and \ref{fig4} maintain the HZN condition, $\frac{dM}{d\rho_{c}}>0$, indicating stable stellar equilibrium. Further, as the HZN condition describes the onset of instability, beyond these configurations, we obtain $\frac{dM}{d\rho_{c}}<0$, i.e. collapsed or unstable stellar structures. Moreover, to substantiate the results, we have tabulated the model parameters at the turning point $(\frac{dM}{d\rho_{c}}=0)$ in Table~\ref{tab6} for the two EoSs employed here. 

\begin{table}[h!]
\caption{\raggedright Tabulation of model parameters at the HZN turning point for (a) BSk24 EoS and (b) SLy4 EoS considering different values of $\beta$.}
\label{tab6}
    \begin{minipage}{0.45\textwidth}
    \centering
    \begin{tabular}{cccc}
	\hline
	\hline
	\multirow{3}{*}{$\beta$} & \multirow{3}{*}{M $\mathrm{(M_{\odot}})$} & \multicolumn{2}{c}{Parameters at turning point} \\ \cline{3-4}
	& & $\rho_{c}\times\,10^{15}$ & $\frac{dM}{d\rho_{c}}$ \\
	& & $(gm/cm^{3})$ & $(Km^{-3})$ \\
	\hline
	0.01 & 2.26 & 2.26 & 5.53$\times10^{-18}$\\
	0.05 & 2.17 & 2.27 & 0.0009 \\
	0.1 & 2.07 & 2.31 & 8.73$\times\,10^{-18}$ \\
	\hline 
	\hline
\end{tabular}
	   \subcaption{}
    \end{minipage}
\hfil
\begin{minipage}{0.45\textwidth}
	\centering
    \begin{tabular}{cccc}
	\hline
	\hline
	\multirow{3}{*}{$\beta$} & \multirow{3}{*}{M $\mathrm{(M_{\odot}})$} & \multicolumn{2}{c}{Parameters at turning point} \\ \cline{3-4}
	&  & $\rho_{c}\times\,10^{15}$ & $\frac{dM}{d\rho_{c}}$ \\
	& & $(gm/cm^{3})$ & $(Km^{-3})$ \\
	\hline
	0.01 & 2.04 & 2.87 & 6.39$\times10^{-18}$\\
	0.05 & 1.96 & 3.41 & -1.17$\times10^{-20}$\\
	0.1 & 1.86 & 3.05 & 1.94$\times10^{-7}$ \\
	\hline 
	\hline
\end{tabular}
    \subcaption{}
    \end{minipage}
\end{table}
\section{Summary and conclusions}

In this work, we have investigated the structural properties of neutron stars in the chameleon scalar–tensor theory of gravity, focusing on the effects of the scalar–matter coupling on stellar equilibrium and on the compatibility of the resulting configurations with observational constraints. To account for uncertainties in the description of dense nuclear matter, we have considered two realistic equations of state, namely BSk24 and SLy4, and numerically solved the modified Tolman–Oppenheimer–Volkoff equations for different values of the coupling parameter $\beta$. Throughout the analysis, the chameleon energy scale and the exponent of the scalar-field potential have been fixed at $\Lambda=10^{-18}M_{\rm P}$ and $n=1$, respectively.

\smallskip

Our results demonstrate that the scalar–matter coupling can produce appreciable modifications to neutron-star structure even when realistic nuclear equations of state are employed. In particular, increasing $\beta$ shifts the mass–radius relations towards lower masses and smaller radii, reducing the maximum mass supported by a given equation of state. For the BSk24 model, the maximum mass decreases from approximately $2.28~M_\odot$ in the general-relativistic limit to $2.07~M_\odot$ for $\beta=0.1$, while the corresponding radius decreases from approximately $11.09~km$ to $10.02~km$. A similar trend is obtained for SLy4, for which the maximum mass decreases from approximately $2.06~M_\odot$ to $1.86~M_\odot$, accompanied by a reduction in radius from approximately $10.00~km$ to $9.06~km$. These results establish that the strength of the chameleon coupling is an important determinant of the macroscopic properties of neutron stars.

\smallskip

The comparison between the two equations of state further reveals that observational constraints on chameleon gravity cannot be considered independently of the underlying nuclear-matter description. The BSk24 equation of state can accommodate the massive pulsars considered in this work over portions of the investigated parameter space, whereas the lower maximum mass predicted by SLy4 imposes more restrictive conditions. In particular, within the configurations examined here, SLy4 cannot reproduce the adopted mass interval of MSP J0740+6620, even in the absence of a significant chameleon-induced reduction. Thus, changing the gravitational sector does not necessarily compensate for limitations associated with a particular equation of state.

\smallskip

We have also examined constraints on $\beta$ obtained from massive-pulsar observations and from the inferred radius of a $1.4~M_\odot$ neutron star. The resulting parameter ranges depend on both the equation of state and the observational benchmark employed. Furthermore, the reference radius of 11.0 km can be reproduced for $\beta=0.126$ with BSk24 and $\beta=0.057$ with SLy4. This illustrates the degeneracy between modifications of gravity and uncertainties in dense nuclear matter: different combinations of the scalar–matter coupling and the equation of state can yield similar stellar observables. Consequently, matching an individual mass or radius measurement is not sufficient to establish the overall viability of a given model, and the different observational constraints must ultimately be assessed together.

\smallskip

The present analysis shows that neutron-star observations provide a useful means of probing the chameleon mechanism in the strong-gravity regime. At the same time, the constraints obtained here should be understood within the adopted potential, the fixed values of $\Lambda$ and $n$, and the two equations of state considered. A more comprehensive investigation should combine mass and radius measurements within a consistent statistical framework, incorporate the relevant observational uncertainties, and explore a wider range of chameleon parameters and nuclear equations of state. An explicit analysis of radial stability would also be valuable for establishing the physical viability of the equilibrium configurations. These extensions would help determine more precisely the extent to which realistic neutron stars can distinguish chameleon gravity from general relativity and constrain screened modifications of gravity.

\section*{Acknowledgments}
DB is thankful to the Department of Science and Technology (DST), Govt. of India, for providing the fellowship vide no:  DST/INSPIRE Fellowship/2021/IF210761. PKC gratefully acknowledges support from IUCAA, Pune, India under Visiting Associateship programme.


\begin{thebibliography}{99}
%
\bibitem{NS1} J.~M.~Lattimer and M.~Prakash,
``The physics of neutron stars,''
Science \textbf{304}, 536-542 (2004)
[arXiv:astro-ph/0405262 [astro-ph]].

\bibitem{NS2} F.~{\"O}zel and P.~Freire,
``Masses, Radii, and the Equation of State of Neutron Stars,''
Ann. Rev. Astron. Astrophys. \textbf{54}, 401-440 (2016)
[arXiv:1603.02698 [astro-ph.HE]].

\bibitem{NS3} S.~Safi-Harb,
``Neutron stars: Observational diversity and evolution,''
J. Phys. Conf. Ser. \textbf{932}, 012005 (2017)
[arXiv:1712.06040 [astro-ph.HE]].

\bibitem{GR} A.~Einstein,
``The foundation of the general theory of relativity.,''
Annalen Phys. \textbf{49}, no.7, 769-822 (1916).

\bibitem{mod1} T.~P.~Sotiriou and V.~Faraoni,
  ``f(R) Theories Of Gravity,''
  Rev.\ Mod.\ Phys.\  {\bf 82} (2010) 451
  [arXiv:0805.1726 [gr-qc]].

\bibitem{mod2} A.~De Felice and S.~Tsujikawa,
  ``f(R) theories,''
  Living Rev.\ Rel.\  {\bf 13} (2010) 3
  [arXiv:1002.4928 [gr-qc]].

\bibitem{mod3} R.~A.~Hennigar, D.~Kubiz{\v{n}}{\'a}k, R.~B.~Mann and C.~Pollack,
``On taking the D {\textrightarrow} 4 limit of Gauss-Bonnet gravity: theory and solutions,''
JHEP \textbf{07}, 027 (2020)
[arXiv:2004.09472 [gr-qc]].

\bibitem{mod4} P.~G.~S.~Fernandes, P.~Carrilho, T.~Clifton and D.~J.~Mulryne,
``Derivation of Regularized Field Equations for the Einstein-Gauss-Bonnet Theory in Four Dimensions,''
Phys. Rev. D \textbf{102}, no.2, 024025 (2020)
[arXiv:2004.08362 [gr-qc]].

\bibitem{mod5} G.~W.~Horndeski,
``Second-order scalar-tensor field equations in a four-dimensional space,''
Int. J. Theor. Phys. \textbf{10}, 363-384 (1974).

\bibitem{mod6} C.~de Rham, G.~Gabadadze and A.~J.~Tolley,
``Resummation of Massive Gravity,''
Phys. Rev. Lett. \textbf{106}, 231101 (2011)
[arXiv:1011.1232 [hep-th]].

\bibitem{mod7} K.~S.~Stelle,
``Renormalization of Higher Derivative Quantum Gravity,''
Phys. Rev. D \textbf{16}, 953-969 (1977).

\bibitem{P1} N.~Yunes, M.~C.~Miller and K.~Yagi,
``Gravitational-wave and X-ray probes of the neutron star equation of state,''
Nature Rev. Phys. \textbf{4}, no.4, 237-246 (2022)
[arXiv:2202.04117 [gr-qc]].

\bibitem{P2} P.~Brax, A.~C.~Davis and R.~Jha,
``Neutron Stars in Screened Modified Gravity: Chameleon vs Dilaton,''
Phys. Rev. D \textbf{95}, no.8, 083514 (2017)
[arXiv:1702.02983 [gr-qc]].

\bibitem{P3} A.~Dima, M.~Bezares and E.~Barausse,
``Dynamical chameleon neutron stars: Stability, radial oscillations, and scalar radiation in spherical symmetry,''
Phys. Rev. D \textbf{104}, no.8, 084017 (2021)
[arXiv:2107.04359 [gr-qc]].

\bibitem{Brans} C.~Brans and R.~H.~Dicke,
``Mach's principle and a relativistic theory of gravitation,''
Phys. Rev. \textbf{124}, 925-935 (1961).

\bibitem{P4} J.~Khoury and A.~Weltman,
``Chameleon fields: Awaiting surprises for tests of gravity in space,''
Phys. Rev. Lett. \textbf{93}, 171104 (2004)
[arXiv:astro-ph/0309300 [astro-ph]].

\bibitem{P5} J.~Khoury and A.~Weltman,
``Chameleon cosmology,''
Phys. Rev. D \textbf{69}, 044026 (2004)
[arXiv:astro-ph/0309411 [astro-ph]].

\bibitem{P6} R.~C.~Tolman,
``Static solutions of Einstein's field equations for spheres of fluid,''
Phys. Rev. \textbf{55} (1939), 364-373.

\bibitem{P7} J.~R.~Oppenheimer and G.~M.~Volkoff,
``On massive neutron cores,''
Phys. Rev. \textbf{55} (1939), 374-381

\bibitem{P8} J.~Bachs-Esteban, I.~Lopes and J.~Rubio,
``Structural Implications of the Chameleon Mechanism on White Dwarfs,''
Universe \textbf{11}, no.7, 237 (2025)
doi:10.3390/universe11070237
[arXiv:2407.04791 [gr-qc]].

\bibitem{P9} G.~Creci, T.~Hinderer and J.~Steinhoff,
``Tidal properties of neutron stars in scalar-tensor theories of gravity,''
Phys. Rev. D \textbf{108}, no.12, 124073 (2023)
[erratum: Phys. Rev. D \textbf{111}, no.8, 089901 (2025)]
[arXiv:2308.11323 [gr-qc]].



\bibitem{Babichev} E. Babichev, D. Langlois, Relativistic stars in $f(R)$ and scalar-tensor theories, Phys. Rev. D {\bf 81}, 124051 (2010), [\url{arXiv:0911.1297 [gr-qc]}].

\bibitem{Bettoni} D. Bettoni, J. Rubio, Quintessential Inflation: A Tale of Emergent and Broken Symmetries, Galaxies {\bf 10}, 22 (2022), [\url{arXiv:2112.11948 [astro-ph.CO]}].

\bibitem{Babichev1} E. Babichev, C. Deffayet, R. Ziour, k-Mouflage gravity, Int. J. Mod. Phys. D {\bf 18}, 2147 (2009), [\url{arXiv:0905.2943 [hep-th]}].

\bibitem{Brax} P. Brax and P. Valageas, The effective field theory of K-mouflage, J. Cosmol. Astropart. Phys. {\bf 01}, 020 (2016), [\url{arXiv:1509.00611 [astro-ph.CO]}].

\bibitem{Vainshtein} A. I. Vainshtein, To the problem of nonvanishing gravitation mass, Phys. Lett. B {\bf 39}, 393 (1972).

\bibitem{Burrage} C. Burrage et al., Massive Galileons and Vainshtein screening, J. Cosmol. Astropart. Phys. {\bf 02}, 050 (2021), [\url{arXiv:2008.01456 [hep-th]}].

\bibitem{Sakstein} J. Sakstein, Stellar oscillations in modified gravity, Phys. Rev. D {\bf 88}, 124013 (2013), [\url{arXiv:1309.0495 [astro-ph.CO]}].

\bibitem{deAguiar} B. F. deAguiar and R. F. Mendes, Highly compact neutron stars and screening mechanisms: Equilibrium and stability, Phys. Rev. D {\bf 102}, 24064 (2020), [\url{arXiv:2006.10080 [gr-qc]}].

 \bibitem{Wang} M. Wang, G. Audi, F. G. Kondev, W. J. Huang, S. Naimi, and X. Xu, The NUBASE2016 evaluation of nuclear properties, Chin. Phys. C {\bf 41}, 030003 (2017), [\url{10.1088/1674-1137/41/3/030001}].

\bibitem{Allard} V. Allard and N. Chamel, $^{1}S_{0}$ pairing gaps, chemical potential and entrainment matrix in superfluid neutron-star cores for the Brussels-Montreal functionals, Universe {\bf 7}, 470 (2021), [\url{arXiv:2203.08778 [nucl-th]}].

\bibitem{Pearson} J. M. Pearson and N. Chamel, Unified equations of state for cold nonaccreting neutron stars with Brussels-Montreal functionals. III. Inclusion of microscopic corrections to pasta phases, Phys. Rev. C {\bf 105}, 015803 (2022), [\url{https://doi.org/10.1103/PhysRevC.105.015803}].

\bibitem{Pearson1} J. M. Pearson, N. Chamel, and A. Y. Potekhin, Unified equations of state for cold nonaccreting neutron stars with Brussels-Montreal functionals. II. Pasta phases in semiclassical approximation, Phys. Rev. C {\bf 101}, 015802 (2020), [\url{arXiv:2001.03876 [astro-ph.HE]}]. 

\bibitem{Pearson2} J. M. Pearson et al., Unified equations of state for cold non-accreting neutron stars with Brussels-Montreal functionals - I. Role of symmetry energy, Mon. Not. R. Astron. Soc. {\bf 481}, 2994 (2018), [\url{arXiv:1903.04981 [astro-ph.HE]}].

\bibitem{Goriely} S. Goriely, N. Chamel, and J. M. Pearson, Further explorations of Skyrme-Hartree-Fock-Bogoliubov mass formulas. XIII. The 2012 atomic mass evaluation and the symmetry coefficient, Phys. Rev. C {\bf 88}, 024308 (2013), [\url{https://doi.org/10.1103/PhysRevC.88.024308}].

\bibitem{Perot} L. Perot, N. Chamel, and A. Sourie, Role of the symmetry energy and the neutron-matter stiffness on the tidal deformability of a neutron star with unified equations of state, Phys. Rev. C {\bf 100}, 035801 (2019), [\url{arXiv:1910.13202 [nucl-th]}].

\bibitem{Xu} Y. Xu, S. Goriely, A. Jorissen, G. L. Chen, and M. Arnould, BRUSsels Nuclear LIBrary (BRUSLIB), Nuclear Astrophysics Compilation of REactions II (NACRE II) and Nuclear NETwork GENerator (NETGEN), Astronomy \& Astrophysics {\bf 549}, A106 (2013), [\url{https://doi.org/10.1051/0004-6361/201220537}].

\bibitem{Welker} A. Welker et al, Binding Energy of $^{79}Cu$: Probing the Structure of the Doubly Magic $^{78}Ni$ from Only One Proton Away, Phys. Rev. Lett. {\bf 119}, 192502 (2017), [\url{https://doi.org/10.1103/PhysRevLett.119.192502}].

\bibitem{Gulminelli} F. Gulminelli and Ad. R. Raduta, Unified treatment of subsaturation stellar matter at zero and finite temperature, Phys. Rev. C {\bf 92}, 055803 (2015), [\url{arXiv:1504.04493[nucl-th]}].

\bibitem{Chabanat} E. Chabanat, Ph.D. thesis, University Claude Bernard Lyon-1, Lyon, France, (1995).

\bibitem{Danielewicz} P. Danielewicz and J. Lee, Symmetry Energy I: Semi-Infinite Matter, Nucl. Phys. A {\bf 818}, 36 (2009), [\url{arXiv:0807.3743[nucl-th]}].

\bibitem{Arzoumanian} Z. Arzoumanian et al., The NANOGrav 11-year Data Set: High-precision Timing of 45 Millisecond Pulsars, Astrophys. J. Suppl. Ser. {\bf 235}, 37 (2018), [\url{arXiv:1801.01837 [astro-ph.HE]}]. 

\bibitem{Cromartie} H. T. Cromartie et al., Relativistic Shapiro delay measurements of an extremely massive millisecond pulsar, Nat. Astron. {\bf 4}, 72 (2020), [\url{arXiv:1904.06759 [astro-ph.HE]}]. 

\bibitem{Antoniadis} J. Antoniadis et al., A Massive Pulsar in a Compact Relativistic Binary, Science {\bf 340}, 448 (2013), [\url{arXiv:1304.6875 [astro-ph.HE]}].

\bibitem{Abbott} B. P. Abbott et al., GW170817: Observation of Gravitational Waves from a Binary Neutron Star Inspiral, Phys. Rev. Lett. {\bf 119}, 161101 (2017), [\url{arXiv:1710.05832 [gr-qc]}].

\bibitem{Capano} C. D. Capano et al., Stringent constraints on neutron-star radii from multimessenger observations and nuclear theory, Nat. Astron. {\bf 4}, 625 (2020), [\url{arXiv:1908.10352 [astro-ph.HE]}].

\bibitem{Abbott1} R. Abbott et al., GW190814: Gravitational Waves from the Coalescence of a 23 Solar Mass Black Hole with a 2.6 Solar Mass Compact Object, Astrophys. J. Lett. {\bf 896}, L44 (2020), [\url{arXiv:2006.12611 [astro-ph.HE]}].

\bibitem{Guver} T. G\"uver, P. Wroblewski, L. Camarota and F. \"Ozel, The Mass and Radius of the Neutron Star in 4U 1820-30. Astrophys. J. {\bf 719}, 1807 (2010), [\url{arXiv:1002.3825 [astro-ph.HE]}].

\bibitem{Guver1} T. G\"uver, F. \"Ozel, A. Cabrera-Lavers and P. Wroblewski, The Distance, Mass, and Radius of the Neutron Star in 4U 1608-52. Astrophys. J. {\bf 712}, 964 (2010), [\url{arXiv:0811.3979 [astro-ph]}].

\bibitem{Abubekerov} M.K. Abubekerov et al., The Mass of the Compact Object in the X-Ray Binary Her X-1/HZ Her. Astron. Rep. {\bf 52}, 379 (2008), [\url{arXiv:1201.5519 [astro-ph.SR]}].

\bibitem{Harrison} B.~K.~Harrison, K.~S.~Thorne, M.~Wakano, and J.~A.~Wheeler, \textit{Gravitation Theory and Gravitational Collapse}, 1965.

\bibitem{Zeldovich} Ya B. Zel'dovich and I.~D.~Novikov, {\it Relativistic Astrophysics, Stars and Relativity} {\bf 1}, (University of Chicago Press, Chicago, 1971).


%
\end{thebibliography}
\end{document}